\documentclass[aps, prd, twocolumn, superscriptaddress, nofootinbib]{revtex4}
\usepackage{graphicx}
\usepackage{dcolumn}
\usepackage{amssymb}
\usepackage{amsmath,amssymb,amsfonts}

\begin{document}

\newcommand{\Eq}[1]{\mbox{Eq. (\ref{eqn:#1})}}
\newcommand{\Fig}[1]{\mbox{Fig. \ref{fig:#1}}}
\newcommand{\Sec}[1]{\mbox{Sec. \ref{sec:#1}}}

\newcommand{\PHI}{\phi}
\newcommand{\PhiN}{\Phi^{\mathrm{N}}}
\newcommand{\vect}[1]{\mathbf{#1}}
\newcommand{\Del}{\nabla}
\newcommand{\unit}[1]{\;\mathrm{#1}}
\newcommand{\x}{\vect{x}}
\newcommand{\ScS}{\scriptstyle}
\newcommand{\ScScS}{\scriptscriptstyle}
\newcommand{\xplus}[1]{\vect{x}\!\ScScS{+}\!\ScS\vect{#1}}
\newcommand{\xminus}[1]{\vect{x}\!\ScScS{-}\!\ScS\vect{#1}}
\newcommand{\diff}{\mathrm{d}}

\newcommand{\be}{\begin{equation}}
\newcommand{\ee}{\end{equation}}
\newcommand{\bea}{\begin{eqnarray}}
\newcommand{\eea}{\end{eqnarray}}
\newcommand{\vu}{{\mathbf u}}
\newcommand{\ve}{{\mathbf e}}


\title{The measure matters}

\newcommand{\addressImperial}{Theoretical Physics, Blackett Laboratory, Imperial College, London, SW7 2BZ, United Kingdom}
\newcommand{\addressRoma}{Dipartimento di Fisica, Universit\`a "La Sapienza"
and Sez. Roma1 INFN, P.le A. Moro 2, 00185 Roma, Italia}

\author{Giovanni Amelino-Camelia}
\affiliation{\addressRoma}
\author{Michele Arzano}
\affiliation{\addressRoma}
\author{Giulia Gubitosi}
\affiliation{\addressRoma}
\author{Jo\~{a}o Magueijo}
\affiliation{\addressImperial}
\affiliation{\addressRoma}

\date{\today}

\begin{abstract}
We adopt a framework where quantum-gravity's 
dynamical dimensional reduction of spacetime at short distances 
is described in terms
of modified dispersion relations. We observe that by subjecting
such models to a momentum-space diffeomorphism
one obtains a ``dual picture'' with unmodified dispersion relations,
but a modified measure of integration over momenta. We then find that the UV
{\it Hausdorff}
dimension of momentum space which can be inferred from this modified integration measure
coincides with the short-distance {\it spectral} dimension 
of spacetime. 
This result sheds  light into why scale-invariant
fluctuations are obtained if the original model for
two UV spectral dimensions is combined with
Einstein gravity. By studying the
properties of the inner product we derive the result that it is only
in 2 energy-momentum dimensions that microphysical vacuum fluctuations
are scale-invariant. This is true ignoring gravity, but then we find that
if Einstein gravity is postulated in the original frame, 
in the dual picture gravity
switches off, since all matter becomes
conformally coupled.
We argue that our findings imply that
the following
concepts are closely connected: scale-invariance of vacuum quantum 
fluctuations, conformal
invariance of the gravitational coupling, UV reduction to spectral dimension
2 in position space and UV reduction to Hausdorff dimension 2 in
energy-momentum space.
\end{abstract}

\keywords{cosmology}
\pacs{}

\maketitle


\section{Introduction}\label{intro2}
In the recent past we have witnessed mounting evidence for the phenomenon of running spectral dimensions in quantum gravity, first found in computer simulations of Causal Dynamical Triangulations (CDT)~\cite{Lollprl}. Examples include
asymptotically safe Quantum Einstein Gravity~\cite{Litim,Reuter}, Ho\v{r}ava-Lifshitz gravity~\cite{HL,HLspec},
spacetime noncommutativity \cite{Alesci:2011cg,Benedetti:2008gu}, spin foams~\cite{Modesto1,Caravelli,Magliaro,Modesto2} and multi-fractional space-times~\cite{Calcagni1,Calcagni2} 
(also see \cite{astrid}). 
Interestingly, not only does it appear that running spectral dimensions are generic in quantum gravity, but it also seems
that most models converge on the prediction that the UV spectral dimension should be 2  (so that we run from 4 spectral dimensions
in the infrared to 2 spectral dimensions in the UV).
Finding observational consequences for running spectral dimensions, and particularly for the case of 2 UV spectral dimensions,
is evidently of paramount importance to research in quantum-gravity. In a recent publication~\cite{dimred} we argued that cosmological fluctuations might be the most fruitful ground for establishing contact with reality.

Central to the work in~\cite{dimred} is the realization that spectral dimensional reduction may be modelled via Planck-scaòle-modified dispersion relations~(MDRs) \cite{HLspec,visser}. Our main result was that a class of modifications of the on-shell relation which produces 2 UV spectral dimensions 
is also associated with 
a scale-invariant spectrum of vacuum fluctuations without appealing to inflation~\cite{Mag}. 
Remarkably, this happens for all equations of state, and for modes inside and outside the horizon. 
The fact that the analysis in~\cite{dimred} establishes a link between 2 UV spectral dimensions (favoured 
on the quantum-gravity side) and a scale-invariant spectrum of vacuum fluctuations (which provides a good first approximation to 
cosmological data) could be very significant and should be understood in depth.
In this paper we investigate further  why scale-invariant
fluctuations are so robustly obtained if the original model for 2 UV 
spectral dimensions
is combined with
Einstein gravity.

The insight we gain is based on the use of ``linearizing'' variables on momentum space for which the original MDR gets rewritten as an unmodified dispersion relation, at the cost of a corresponding modification of the measure of integration over momenta. We essentially adopt  a change of the units used for measuring energy and momentum~\cite{gacdsr,gacreview,MagleePRL,leedsrPRD,DSRposition} such that    the description of the novel Planck-scale effects is shifted from the dispersion relation to the integration measure. This equivalent reformulation of the theory is less convenient
than the original one in most cases, but may shed light on some particular aspects of the theory.
 A rather striking example of the latter possibility is uncovered in this paper.

We already obtained some insight into the structures underlying the findings in Ref.~\cite{dimred} in a follow up study,
reported in~\cite{rainbowred}. In that paper we changed
the unit of time, thereby disformally transforming to a ``rainbow frame''~\cite{rainbowDSR}. This  allowed us to expose special
properties of gravity in the UV regime of the relevant models.
 The change of units of  momentum to be derived
in Section~\ref{cute} leads us to the remarkable realization that  in the UV limit the {\it Hausdorff} dimension of energy-momentum space (in the dual picture) coincides with 
the {\it spectral} dimension of spacetime. 
Given the apparent robustness of this mechanism we conjecture that 
this result should hold in any quantum-gravity
picture leading to UV spectral dimensional reduction, even though our analysis remains confined to the specific picture
on which  Ref.~\cite{dimred} is based.

The rest of this paper puts to work the picture emerging from Section~\ref{cute}, deriving from it insights into the mechanism producing
pervasive scale invariance of density fluctuations in~\cite{dimred}.
In Section~\ref{vacflucts} we first ignore the effects of cosmological expansion, and examine how the modified measure 
of integration on momentum space
affects the vacuum fluctuations. We find that the form of the measure bears directly upon the definition of scale invariant fluctuations (Subsection~\ref{scinvdef}); however it does not change the Hilbert space's inner product, essential for defining creation and annihilation operators, and so the amplitude of the vacuum fluctuations (Subsection~\ref{inner}). Putting the two effects together we find the notable result that {\it it is only with 2 energy-momentum dimensions that microphysical vacuum quantum fluctuations are scale-invariant} (Subsection~\ref{vacflucts1}).

In Sections~\ref{uvgrav} and~\ref{irgrav} we include gravity in our considerations, resolving first a number of technical issues. Since we want to work with comoving wavenumbers (and not time dependent ones) in Section~\ref{uvgrav} we have to adopt ``linearizing'' units following 2 separate steps. We first ``linearize'' ignoring the effects of expansion on the wavelengths: this leads to the same Hausdorff reduction as in Minkowski space-time, but leaves us with a time-dependent (but $k$-independent) speed of light (so that linearization has not been fully achieved). The latter is then removed by going to the ``matter frame'', which, unlike in~\cite{rainbowred}, is {\it not} a rainbow frame.

Just as in~\cite{rainbowred} we discover that after these operations are performed, all matter becomes conformally coupled, and so impervious to the effects of gravity in a flat Friedmann model. Consequently there are no horizons, and the modes are always in the regime of Section~\ref{vacflucts}, where expansion was ignored. Thus, we have pervasive scale-invariance even in the presence of gravity in the original frame.
The core of our conclusions on gravity is in Section~\ref{uvgrav}, but we must  still explain why scale-invariance is preserved as the modes go from UV to IR due to the stretching effects of expansion. We do this in Section~\ref{irgrav}.
Finally, in Section~\ref{concs} we summarize our results and speculate on possible extensions, specifically how strict scale-invariance might be broken.
We also observe that our results can be viewed as a first attempt at evaluating the impact of the measure
in momentum space on cosmological vacuum quantum fluctuations, 
in a special setting where we can rely on the guidance of a  
dual picture studied comprehensively 
in Refs.\cite{dimred,rainbowred}.

\section{The dual picture of dimensional reduction}\label{cute}
Dimensional reduction may be modelled via a very specific set of
modified dispersion
relations. In general MDRs can be reinterpreted as
a deformation of the measure of momentum space
by transforming to ``linearizing'' units. In this Section we
perform this exercise
with the particular MDR associated with dimensional reduction.
It is well known~\cite{HLspec,visser} that theories with
\be\label{ddr1}
E^2=p^2(1+(\lambda p)^{2\gamma})
\ee
produce running of the spectral dimension from $D+1$
in the IR to
\be\label{ds}
d_S=1+\frac{D}{1+\gamma},
\ee
in the UV. Here $D$ is the number of spatial dimensions of the original
spacetime (taken to be 3 for most of this paper, but left general in this
Section). A possible linearizing variable for (\ref{ddr1})
is the spatial momentum variable:
\be\label{newp}
\tilde{\vect{p}}=\vect{p}\sqrt{1+(\lambda p)^{2\gamma}},
\ee
(where $p=|{\vect{p}}|$)
in terms of which
\be
E^2={\tilde p}^2\, .
\ee
Its use shifts non-trivial effects to the momentum measure, taken to be
undeformed in the original variable $\vect{p}$. Using polar coordinates in
momentum space we find that for $\lambda p\gg 1$ the radial measure
becomes:
\be
p^{D-1}dp \propto {\tilde p}^{\frac{D-1-\gamma}{1+\gamma}}d{\tilde p}
\ee
corresponding to an energy-momentum space with UV Hausdorff dimension:
\be\label{dk}
d_{\tilde p}=2+\frac{D-1-\gamma}{1+\gamma},
\ee
so that
\be
d_S=d_{\tilde p}\, .
\label{wow}
\ee
This dimensional reduction is isotropic, and the angular variables present in the original measure get integrated into a multiplicative constant.

The result (\ref{wow}) is rather nontrivial since it establishes a link between two very distinct notions: the UV {\it spectral} dimension  of {\it spacetime}, $d_S$, and the {\it Hausdorff dimension} of the energy-momentum space, $d_{\tilde p}$. The former is presently understood as a purely formal concept, characterizing a fictitious diffusion process, taking place in a fictional, and hard to interpret diffusion time. In contrast the latter is the {\it Hausdorff dimension} of the energy-momentum space (in units where the measure carries all the non-trivial deformation),  which is of transparent physical interpretation. 

\begin{figure}[h]
\begin{center}
\scalebox{0.7}{\includegraphics{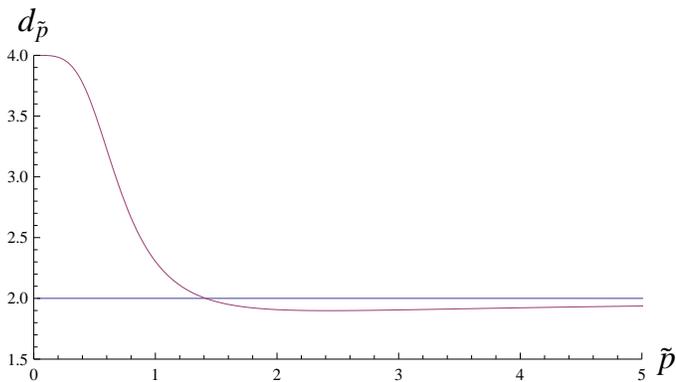}}
\caption{\label{MDR26} Purple line: running the Hausdorff dimension in momentum space for quadratic plus sextic dispersion relation (eq. (\ref{ddr1}) with $\gamma=2$ and $\lambda=1$). The blue line gives a reference for the UV limit of the spectral dimension derived from the same dispersion relation.}
\end{center}
\end{figure}

\begin{figure}[h]
\begin{center}
\scalebox{0.7}{\includegraphics{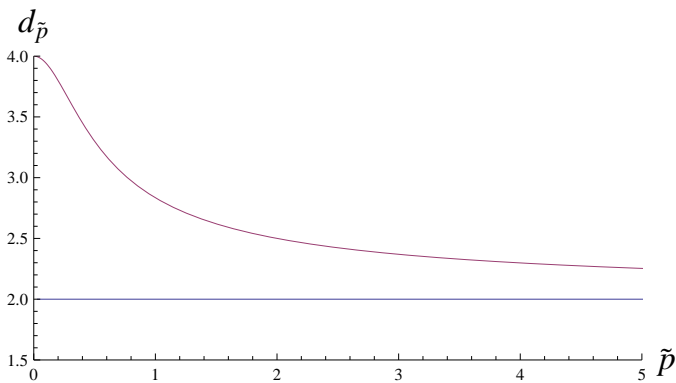}}
\caption{\label{MDR246}Purple line: running the Hausdorff dimension in momentum space for quadratic plus quartic plus sextic dispersion relation (eq. (\ref{ddr1b}) with $\gamma=2$ and $\lambda=1$).The blue line gives a reference for the UV limit of the spectral dimension derived from the same dispersion relation.}
\end{center}
\end{figure}

Notice also that the result we have derived is only valid asymptotically ($\lambda p\gg1$) and for the specific power-law MDRs we are considering. In no way can the full transition function $d_S(s)$ and $d_{\tilde p}({\tilde p})$ be easily mapped into each other. Indeed they can be qualitatively very different. In Fig.~\ref{MDR26} we have plotted $d_{\tilde p}({\tilde p})$ for the MDR (\ref{ddr1}) with $\gamma=2$. We defined this by computing the momentum measure $d\mu({\tilde p})=\mu({\tilde p})d\tilde p$ and then evaluating:
\be
d_{\tilde p}({\tilde p})=2+ \frac{d\log\mu}{d\log \tilde p}.
\ee
We then did the same for the alternative MDR:
\be\label{ddr1b}
E^2=p^2(1+(\lambda p)^{\gamma})^2
\ee
(see Fig.~\ref{MDR246}).
The running of the spectral dimension $d_S(s)$ associated with these dispersion relations was already presented in~\cite{dimred}. We note that the phenomenon of overshooting the asymptotic value ($d_{\tilde p}$ goes below 2 before settling on 2; cf. Fig.~\ref{MDR26}) for the first MDR was not observed in terms of $d_S(s)$.

We also note that regarding the nitty-gritty details of the transition, the dual picture is usually complicated for simple MDR, and vice-versa. It is hard to find an analytical expression for the measure $\mu(\tilde{p})$ associated with (\ref{ddr1}).  Likewise, we can write simple expressions with the correct UV and IR asymptotic behaviour for the measure, such as
\be
d \mu(p)=d p \frac{p^2}{1+(\lambda p)^2}
\ee
but these are not associated with simple MDR, except asymptotically.

\section{Vacuum fluctuations under a non-trivial measure}
\label{vacflucts}
We now examine vacuum fluctuations under a non-trivial measure, ignoring first the effects of expansion. Nonetheless, with a view to generalizing our results, we shall use the curvature variable $\zeta$ employed to quantify fluctuations in expanding universes. We shall also drop the tilde over our momentum variables (and assume that we are already in the dual basis, i.e. we have a deformed measure and undeformed MDR), and replace $p$ by $k$ (the comoving wavenumber used in expanding universes), since the two are equal without expansion.

We will be concerned with the power spectrum of $\zeta$, $P_\zeta(k)$,
defined from:
\be\label{powersp}
{\langle \zeta ({\mathbf k}) \zeta^\star ({\mathbf k}')\rangle}
=\delta({\mathbf k}- {\mathbf k}')P_\zeta (k)
\ee
where $\zeta ({\mathbf k})$ is the Fourier component of $\zeta(\vect{x})$ based on the momentum measure we are working with:
\be\label{fourier}
\zeta(\vect{x})=\int d\mu(\vect{k})\zeta(\vect{k})e^{i\vect{k}\cdot \vect{x}}.
\ee
The delta function is defined with respect to the same measure, i.e.:
\be\label{delta}
\int d\mu(\vect{k})\delta({\mathbf k})=1\, ,
\ee
and to recall this fact we will sometimes label it $\delta_\mu(\vect{k})$.

The ``average'' bracket in (\ref{powersp}) is loosely defined. It may be an ensemble average (observational cosmology), a vacuum expectation value (as in Section~\ref{vacflucts1}), or even a thermal average, if appropriate. The reason why all these averages are identified is beyond the scope of this paper.

\subsection{Scale-invariance under a non-trivial measure}\label{scinvdef}
It is immediately obvious that the definition of scale-invariance depends
on the measure, specifically when the measure implies a change in the
dimensionality of momentum space.
This is easy to see based on dimensional arguments similar to those used by Zeldovich in his initial proposal for scale-invariance.

Let us assume that $D_{k}$ is the number of spatial momentum dimensions (so that in the example of Section~\ref{cute} we have $d_{\tilde p}\rightarrow D_k+1$, and so  $D_k=1$ for $\gamma=2$). In Section~\ref{cute} we have defined the measure to have the same unit-dimensions as it runs from IR to UV, and this is enforced by the dimensional parameter $\lambda$ (for example for $\gamma=2$ we have $d\mu=dk/\lambda^2$ in the UV, and although $D_k=1$ we still have $[d\mu]=1/L^3$). But if we focus on the UV regime, where dimensional reduction has taken place, this makes little sense, and we should use a measure with the same unit-dimensions as that of the space itself.  We should therefore impose the replacement:
\be
d\mu=\frac{k^{D_k-1}}{\lambda ^{D-D_k}}dk\rightarrow
k^{D_k-1} dk\; ,
\ee
so that the measure's unit-dimensions match the UV dimensionality of the space. Concomitant with this replacement we then have:
\bea
\zeta(\vect{k})&\rightarrow &\frac{\zeta(\vect{k})}{\lambda ^{D-D_k}}\\
\delta^D(\vect{k})&\rightarrow &\delta^{D_k}(\vect{k}).
\eea
Once this is done,
given that $\zeta$ is dimensionless in position
space (i.e. as $\zeta(\vect{x})$), we can infer that:
\bea
\left[\zeta(\vect{k})\right]&=&L^{D_k}\\
\left[\delta^{D_k} (\vect{k})\right]&=&L^{D_k}\\
\left[P_\zeta(k))\right]&=&L^{D_k}.
\eea
Consequently, the ``dimensionless'' power
spectrum should be defined as:
\be
{\cal P}_\zeta (k)=k^{D_k} P_\zeta (k)
\ee
and this is the quantity that should be independent from $k$ for a scale-invariant spectrum.

In cosmology $\zeta$ is often used to denote its r.m.s., i.e. the square root of its power spectrum. Then, for a scale-invariant spectrum, $\zeta$ should be:
\be
\zeta=\frac{A}{k^{D_k/2}}
\ee
i.e.
\be
\zeta\sim k^{-3/2}
\ee
in 3 spatial dimensions, as is well known. However, if $D_k=1$ (corresponding to $\gamma=2$), this becomes instead:
\be
\zeta\sim k^{-1/2},
\ee
which should immediately raise a warning flag.



\subsection{The inner product}\label{inner}
The inner product normalization is essential in the computation of the vacuum fluctuations. In this subsection we examine how deforming the measure in momentum space interacts with the inner product. Since we are ignoring expansion we can set up the field theory directly in terms of $\zeta$ (i.e. the auxiliary variable $v$ is the same as $\zeta$). The second order action is:
\be
S_2=\int d\mu(\vect{k})\, d\eta \left[\zeta'^2 + k^2 \zeta^2\right]
\ee
with equations of motion:
\be
\zeta''+ k^2 \zeta=0.
\ee
A set of plane wave solutions is given by
\be
\zeta_{\vect{k}}({\bf x}, \eta) \propto e^{i \vect{k}\cdot \vect{x}} \zeta_{\vect{k}}(\eta)\,,
\ee
whose Fourier modes read
\be
\zeta_{{\bf k}}({\bf p},\eta)\propto\delta_{\mu}({\bf k}-{\bf p})\, \zeta_{{\bf k}}(\eta)\, ,
\ee
where the $\propto$ sign signifies we have yet to fix the normalization.
We want these plane waves to form an orthogonal basis with respect to the inner product
\be\label{innerprod}
(\zeta_{{\bf k}}, \zeta_{{\bf k}'})=i \int d\mu({\bf p})\,\,  \delta_{\mu}({\bf k}+{\bf p})\, \delta_{\mu}({\bf k}'-{\bf p})\,  (\zeta^*_{{\bf k}}(\eta) \overleftrightarrow{\partial_\eta}\, \zeta_{{\bf k}'}(\eta))\, .
\ee
Note that in our linearizing units the dispersion relations are undeformed, and so this product is obviously time-independent. However,
it can be checked that even when we have MDR, Eq.~(\ref{innerprod}) is a conserved inner product as long as the theory is defined as a higher order field theory with higher than second order spatial derivatives only.

We require a normalization for our modes such that
\be
(\zeta_{\vect{k}}, \zeta_{\vect{k}'}) = \delta_\mu(\vect{k}-\vect{k}')\, .
\ee
Therefore we can conclude that,
{\it regardless of the measure on spatial momenta}, the normalized wave modes are given by
\be
\zeta_{\vect{k}}({\bf p},\eta) = \frac{1}{\sqrt{2\omega_{\vect{k}}}} \delta_{\mu}({\bf k}-{\bf p})\, e^{-i\omega_{\vect{k}} \eta}\, .
\ee
with $\omega_{\vect{k}}=k$ in our case. This fixes the spectrum, as we now show.

\subsection{Scale-invariance for  $\gamma=2$}\label{vacflucts1}
The spectrum of perturbation is determined by the vacuum fluctuations of the quantum field.   As customary in quantizing a classical system one promotes classical observables (i.e. functions on phase space) to operators, and their Poisson brackets to commutators.  In our case the classical phase space is the space of solutions of the equations of motions which is spanned by our mode functions $\zeta_{{\bf k}}$.  The Poisson bracket is defined for functions on phase space.  The inner product above gives us a natural function on phase space namely $(\zeta, \cdot)$ i.e. the functional which associates to each solution $\phi$ the complex number $(\zeta, \phi)$.  The classical Poisson bracket is then given by~\cite{Wald:1995yp}
\be\label{Poiss1}
\{(\zeta_1, \cdot), (\zeta_2, \cdot)\}=(\zeta_1, \zeta_2).
\ee
On our mode functions $\zeta_{{\bf k}}$ the Poisson bracket above can be written in the more familiar form
\be\label{Poiss2}
\{\zeta_{{\bf k}}, \pi_{{\bf k}}\} = \delta^3({\bf k}+{\bf k}')
\ee
The most direct way of quantizing our field is to think of $\zeta_{{\bf k}}(\eta)$ as ``position coordinates" of a time-dependent harmonic oscillator. In the Schroedinger picture one promotes the ``coordinates" to operators
\be
\hat{\zeta}_{{\bf k}} = \frac{1}{\sqrt{2\omega_{{\bf k}}}} (a_{{\bf k}}+ a^\dagger_{-{\bf k}})
\ee
while the ``momentum" operators will be the quantized version of the classical ones
\be
\hat{\pi}_{{\bf k}} = -i\sqrt{\frac{\omega_{{\bf k}}}{2}} (a_{{\bf k}} - a^\dagger_{-{\bf k}})\, .
\ee
The commutators of these operators are obtained by quantizing the classical Poisson brackets (\ref{Poiss2}).  The normalization which makes the mode functions orthonormal determines the normalization of the ``position" and ``coordinate" operators and ensures that from
\be
[\hat{\zeta}_{{\bf k}}, \hat{\pi}_{{\bf k}'}] = i\delta^3({\bf k}+{\bf k}')
\ee
one gets
\be
[a_{{\bf k}}, a^{\dagger}_{{\bf k}'}]= \delta^3({\bf k}-{\bf k}')\,.
\ee
The main point is that {\it such normalization does not depend on a change of measure for spatial momenta}.  Finally to get the spectrum of fluctuations one has to go to the Heisenberg picture where
\be
\hat{\zeta}_{{\bf k}} = \frac{1}{\sqrt{2\omega_{{\bf k}}}} (a_{{\bf k}} e^{i\omega_{{\bf k}}\eta}+ a^\dagger_{-{\bf k}}e^{-i\omega_{{\bf k}}\eta})\,.
\ee
We therefore have that 
\be
{\langle 0| \zeta_{\mathbf k}\, \zeta^\star_{{\mathbf k}'}|0\rangle}
=\delta_{\mu}({\mathbf k}- {\mathbf k}')|\zeta_{\vect{k}}(x)|^2
\ee
so that:
\be
P_\zeta(k)=\frac{1}{2\omega_{\vect{k}}}
\ee
The important conclusion is that this result cares about the dispersion relations but not about the measure. Therefore in our dual units (with $\omega_{\vect{k}}=k$) we have:
\be
P_\zeta(k)=\frac{1}{2k}
\ee
for all values of $\gamma$ (and so for all values of momentum dimensions $D_k$).
However, as we explained in subsection~\ref{scinvdef}, this is actually the definition of scale invariance when $D_k=1$. 

Therefore we arrive at the very
interesting result that {\it only} for $D_k=1$ (corresponding to $d_{\tilde p}=2$
and $\gamma=2$) are the vacuum fluctuations scale invariant.

\subsection{An alternative derivation}
An alternative way to derive this result, closer to field theory but more unfamiliar to cosmologists, consists of requiring that two-point function in position space be independent of the separation.

Let us look at the vacuum fluctuations of our quantum field at a given time and at two spatial points separated by a distance $L= |\vect{x}-\vect{y}|$.  If the momentum measure is undeformed, we have
\be
\langle 0|\hat{\zeta}(\vect{x},\eta)\,\hat{\zeta}(\vect{y},\eta) |0\rangle \sim \int_0^{\infty} k^2 dk\, |\zeta_{\vect{k}}|^2\, \frac{\sin(kL)}{kL} \sim\, k^3\, |\zeta_{\vect{k}}|^2
\ee
where we used the usual relation $\langle 0|a_{\vect{k}} a^{\dagger}_{\vect{k}'} |0\rangle = \delta^3(\vect{k}-\vect{k}')$ and the fact that the main contribution to the integral comes from values of the spatial momentum for which $kL\sim 1$ (so in the last equality one has $k\sim 1/L$).
We see that a scale invariant spectrum of fluctuations can be achieved when $|\zeta_{\vect{k}}|^2 \sim \omega_{\vect k}^{-1}=k^{-3}$.  This is true under the MDR $\omega_{\vect{k}}\sim k^3$, as we know.

In our ``linearized'' picture the dispersion relation is {\it undeformed} (and thus $\omega_{{\bf k}} = k$) but the Hausdorff dimension of momentum space runs to 2 so that spatial momenta become one-dimensional.  In this case the Dirac deltas in the relations above become one-dimensional together with the integration measures on momentum space.   We see that the dimensional running to two in momentum space is very special indeed since, as it also leads to a scale invariant spectrum, but for different reasons. Specifically:
\bea
\langle 0|\hat{\zeta}(\vect{x},\eta)\,\hat{\zeta}(\vect{y},\eta) |0\rangle &\sim& \int_0^{\infty} dk\, |\zeta_{{\bf k}}|^2\, \frac{\sin(kL)}{kL} \nonumber\\
&\sim& k\, |\zeta_{\vect{k}}|^2 \sim k\, k^{-1}\, ,
\eea
and once more scale-invariance is achieved. The two derivations presented are exactly equivalent, but different readers might find one easier to understand than the other.

\section{UV gravity in the dual frame}\label{uvgrav}
If we now take expansion into account we end up with a hybrid between the deformed measure picture we derived for Minkowski spacetime and the bimetric (time-dependent) varying speed of sound scenario~\cite{csdot,bim,ngbim}. Note that in cosmology one works with {\it comoving} wave vectors, which are fixed labels for modes being stretched by expansion, and can be seen as the conserved charges associated with spatial translational invariance or homogeneity. Therefore in the original frame (where (\ref{ddr1}) is valid) the speed of light for each mode is time-dependent, since its physical wavelength (entering the MDR) is stretched by expansion. An originally purely energy-dependent speed of light thus
becomes time-dependent,  due to the effects of expansion. In the  UV:
\be\label{cofE}
c=\frac{E}{p}\propto (\lambda p)^\gamma=
{\left(\frac{\lambda k}{a}\right)}^\gamma\; .
\ee
This has the unfortunate effect of making the ``linearizing'' variable (rendering $c=1$) time-dependent, i.e.:
\be
{\tilde k}={\tilde k}(\eta)=c(k,\eta)k.
\ee
Cosmological calculations are invariably carried out with time-independent $k$, and such a time-dependent ``comoving'' wave-vectors would require a significant revision of the formalism.

The solution to this consists of ``linearizing'' under expansion in two steps. First we redefine momenta ignoring the stretching effects of expansion, just like we did in Section~\ref{cute}, but using comoving wavevectors. As we will see this still leaves us with a varying speed of light, but it turns out that in the UV this is now purely time-dependent. We can then move to a fixed-$c$ frame by changing the time unit, following~\cite{piazza,ngbim}. But unlike in~\cite{rainbowred} the new frame is {\it not} $k$-dependent and therefore it is not a rainbow frame.

More concretely, we start with
the quadratic action in the original frame:
\be\label{s2}
S_2=\int d\eta\, d^3k\,a^2 \left[\zeta'^2 + c^2 k^2 \zeta^2\right]\; .
\ee
We then apply the transformations in Section~\ref{cute} to
comoving momenta, ignoring expansion, defining:
\be\label{newk}
{\tilde k}=k\sqrt{1+(\lambda k)^{2\gamma}}.
\ee
This mimics (\ref{newp}) and  in the UV it becomes ${\tilde k}\approx k(\lambda k)^{\gamma}$. As in Section~\ref{cute} this transformation leads to a dimensional reduction of momentum space. The speed of light also becomes $k$-independent, just as in that Section. However we find that we have not removed the time-dependence in $c$, imparted by  expansion. Indeed the action is now:
\be\label{s2a}
S_2=\int d\eta\, d{\tilde k}{\tilde k}^{D_k-1} \, a^2 \left[\zeta'^2 + \frac{{\tilde k}^2}{a^{2\gamma}} \zeta^2\right]\; .
\ee
and we note the presence of a time-dependent (but not momentum dependent) speed of light, with $c\propto a^{-\gamma}$. In this expression $d_k=D_k+1$, as given in Section~\ref{cute}, Eq.~(\ref{dk}) (so that $D_k=1$ for $\gamma=2$).

The residual varying $c$ can be removed following the trick in~\cite{piazza,ngbim} of redefining the time unit:
\be
d\tau=a^{-\gamma}d\eta.
\ee
However we are now applying this trick in its strict sense, 
rather than the adaptation
described in~\cite{rainbowred}. In the latter the new time was $k$-dependent
and so we were transforming to the ``rainbow frame'', where $c=1$ even without
a deformation of the measure. Not so here, where the new frame is
$k$-indepedent.

With respect to $\tau$ we finally have:
\be
S_2=\int d\tau\, d{\tilde k}{\tilde k}^{D_k-1} \, z^2 \left[\dot \zeta^2 + {\tilde k}^2 \zeta^2\right]
\ee
where the dot denotes derivative with respect to $\tau$ and
\be
z=a^{1-\frac{\gamma}{2}}.
\ee
This leads to the usual
\be\label{veq} \ddot v +\left[{\tilde k}^2 -\frac{\ddot z}{z}\right]v=0, \ee
with $\zeta=-v/z$, where $z$ controls the effects of expansion.
As in~\cite{rainbowred} we find that a miracle happens when $\gamma=2$,
describing a reduction from 4 to 2 dimensions. We find that:
\be
z=1
\ee
so that the effects of expansion (and Jeans' instability) switch off altogether.
The action and equations of motion become the same as in the Sections where
we ignored expansion. Expansion drops out of the picture.

This phenomenon was already  encountered in~\cite{rainbowred}. As we explained there, this is what happens in the standard theory for radiation (i.e. standard, undeformed radiation) as a result of its conformal invariance. Then $a\propto \eta$ and so the mass term vanishes:
\be
\frac{z''}{z}=\frac{a''}{a}=0
\ee
signalling the conformal invariance of radiation. But for
$\gamma=2$ this happens for
all equations of state. Everything is conformally
invariant  and so fails to feel gravity in a flat Friedmann model
(which is conformal to Minkowski space-time). The results in Sections~\ref{cute} and~\ref{vacflucts} are therefore directly applicable, namely the conclusion that the fluctuations are scale-invariant.

\section{Spectrum in the IR  phase}\label{irgrav}
Obviously a question remains to be answered. As the modes go from UV to IR due to expansion, the spatial momentum space gains two dimensions and becomes 3D, and gravity switches on and becomes ``Einstein'', so that there are now horizons and the relevant modes find themselves outside the horizon. The question, then, is why is it that the spectrum remains scale-invariant with regards to the new, IR, number of momentum dimensions?

In a general set up, this should happen because the gravity theory
is required to imply:
\be
\frac{d}{d\eta}(k^3(\tilde k)\zeta^2)=0,
\ee
i.e. the dimensionless spectrum (adjusted to a varying dimension) must be  a conserved quantity. This could follow, in a general setting, from the Casimir operators of the theory,  or, in a Hamiltonian setting, from the theory's first integrals. In our specific case, however, this unique property can be derived by a simple adaptation of the argument in Section~\ref{uvgrav}.

We can still apply (\ref{newk}) to the action (\ref{s2}) even if we
are not in the UV regime (i.e. Eq.(\ref{cofE}) is not valid).
Then, instead of (\ref{s2a}), we have:
\be
S_2=\int d\eta\, d{\tilde k}{\tilde k}^{D_k-1} \, a^2 \left[\zeta'^2 + {\tilde c}^2{\tilde k}^2 \zeta^2\right]\; ,
\ee
with
\be
{\tilde c}(k,\eta)=\frac{1+\left(\frac{\lambda k}{a}\right)^\gamma}{1+(\lambda k)^\gamma}.
\ee
Obviously $c$ is both time and $k$ dependent unless $\lambda k\gg 1$
and $\lambda k/a\gg 1$. Therefore as the mode goes from from UV to IR
the matter frame
transformation is $k$-dependent, and so associated with a rainbow metric,
as in~\cite{rainbowred}.

Note that since $k$ is time-independent, the assumption $\lambda k\gg 1$
must be valid always, so we can simplify:
\be
{\tilde c}(k,\eta)=\frac{1}{(\lambda k)^\gamma}+ \frac{1}{a^\gamma}.
\ee
The UV limit corresponds to the second term dominating the first.
As in~\cite{rainbowred}, the new time variable should be defined as
\be
\tau=\int \tilde c(k,\eta)\, d\eta
\ee
in terms of which the action is
\be
S_2=\int d\tau\, d{\tilde k}{\tilde k}^{D_k-1} \, z^2 \left[\dot \zeta^2 + {\tilde k}^2 \zeta^2\right]\; .
\ee
with
\be
z=a\sqrt{\tilde c}.
\ee
This is all we need to derive the spectrum in the IR.  The equation of motion is still (\ref{veq}) but now $z$ is only $1$ asymptotically, in the UV, becoming the usual $z\propto a$ in the IR. At all times, for modes outside the horizon (i.e. dominated by the $\ddot z/z$ term), we have the growing mode solution:
\be
v\propto z
\ee
and therefore $\zeta=- v/z$ is a constant. This is true even as the modes go from UV to IR. Since in the UV we have $\zeta=1/\sqrt{\tilde k}$, this must be true in the IR, too.

As the mode is stretched deeper into the IR, we find ourselves with the standard theory, but written in terms of physically irrelevant variables. This had to be the case, since the transformation we applied to  $k$ is time-independent, and therefore cannot know that the physical momentum went from UV to IR. Specifically we have $c\approx 1$ but we are still using $\tilde k=\lambda^2k^3$. Likewise we are using a rainbow time variable ($d\tau=d\eta/(\lambda k)^2$) for no good reason. We should therefore transform back to $k$ and $\eta$ variables, to recover action (\ref{s2}), with $c\approx 1$. In the process we find that:
\be
\zeta\sim \frac{1}{\sqrt{\tilde k}}=\frac{1}{\lambda k^{3/2}}
\ee
i.e. 3D scale-invariance.

\section{Conclusion: the measure matters}\label{concs}
In this paper we made a first attempt at evaluating the impact of the measure
in momentum space on vacuum quantum fluctuations. We did so with a safety net:
we examined a theory with a deformed measure which is the dual picture of 
a theory which we have studied comprehensively (\cite{dimred,rainbowred}) 
and for which we ``know the result''.
By fully transferring the non-trivial effects from the dispersion relations
to the measure, in Section~\ref{cute} we derived our first notable result. 
We considered the case of spectral dimensional 
reduction as modelled with MDRs and examined the dual picture, where the measure
is deformed, but not the dispersion relations. 
We found that {\it the UV Hausdorff
dimension of momentum space which can be inferred from the 
modified integration measure coincides with the short-distance 
spectral dimension of spacetime}. The two notions are rather different, 
so this is a highly non-trivial statement.

We then derived a number of important results concerning fluctuations,
focusing on the case leading to strict scale-invariance and 2 UV dimensions
(but our arguments can be adapted to more general cases). 
In Section~\ref{vacflucts} we 
concluded that, if the dispersion relations are unmodified, {\it it is only with 2 energy-momentum dimensions that vacuum quantum fluctuations are scale-invariant}. The derivation used the essential fact that the inner product with which
the Hilbert space is endowed does not care about the deformation in the measure.
This result applies to modes inside the horizon, or ignoring gravity.  
However, in Section~\ref{uvgrav} we concluded further that {\it in the dual
frame gravity switches off (or all matter is conformally coupled) if
Einstein gravity is assumed in the original frame, and if we run to 2 UV
dimensions}. Thus the arguments we developed whilst ignoring expansion are
directly applicable even if the universe is expanding.  
This explains why we find ubiquitous scale-invariance without
inflation (appealing instead to a 
varying speed of light~\cite{Albrecht,Moffat,vslreview} solution),
for modes inside and outside the horizon, and for all 
equations of state~\cite{dimred,Mag}. This pervasive scale-invariance
is a direct implication of universal gravitational conformal coupling, and
the scale-invariance of vacuum fluctuations when we run to 2 UV dimensions.

From the derivations in this paper we can infer straightforward extensions
of our results.
Following the arguments in Section~\ref{vacflucts} and~\ref{uvgrav} 
we see that
the general condition for scale-invariance is:
\be\label{mastercond}
k\mu(k)= A^2 \omega(k)
\ee
where $A$ is the spectrum's amplitude and $d\mu=\mu(k)dk$ is the 
spatial momentum measure.
The key  assumptions are:
\begin{itemize}
\item The inner product is blind to the measure (i.e. the assumptions used in Section~\ref{vacflucts} are valid).
\item Gravity is ``innocuous'' (i.e. we have Einstein gravity in some frame,
leading to conformal coupling in another).
\end{itemize}
Condition (\ref{mastercond}) is applicable for general theories in
which it is natural to have a deformed measure as well as  MDR, such as DSR \cite{gacdsr,gacreview,kowadsr,MagleePRL,leedsrPRD}.
But it is also true for dual, equivalent pictures of the same theory. For example,
in~\cite{dimred} the condition is satisfied for $\gamma=2$ because in the UV:
\bea
\mu(k)&=& k^2\\
\omega(k)&=&\lambda^2 k^3
\eea
whereas here it is satisfied because, also in the UV, we have:
\bea
\mu(k)&=& \frac{1}{\lambda^2}\\
\omega(k)&=& k.
\eea

There are several possible interpretations for what we have found, and for how things might change within a deeper analysis. This could be related to the fact that the observed spectrum is not exactly scale-invariant. One intriguing possibility is what might be called ``Polyakov's revenge''.  We found that the classical field theory has ``emergence of scale invariance'' in the UV because in the UV it turns effectively into a 2D theory. This is significant, but we must remember that the ingredients of the quantum version of a theory are the classical theory {\it plus} a measure of integration over Feynman paths of the fields. The Polyakov anomaly consists of the fact that a certain 2D classical theory has more  symmetries than one can implement in the path-integral measure. Therefore we may expect that if we fully quantize the theory we will find that it does become effectively 2D but it is not exactly scale invariant, even in the UV. Could this be the origin of the observed small departures from scale-invariance?

\section*{Acknowledgments}
GAC, MA, and GG were supported in part by the John Templeton Foundation.  The work of MA was also supported by the EU Marie Curie Actions through
a Career Integration Grant.
JM was funded by STFC through a consolidated grant and by an International Exchange Grant from the Royal Society.

\bibliography{refsMEAS}

\end{document}